\documentclass[aps,pre,twocolumn,groupedaddress,showpacs]{revtex4-1}

\usepackage{amsmath}
\usepackage{amssymb}

\usepackage{graphicx}

\newcommand{\bs} {\boldsymbol}

\begin{document}
\newcommand{\ud}{{\mathrm d}}
\newcommand{\sech}{\mathrm{sech}}

\title{Thermal equilibrium in Einstein's elevator}



\author{Bernardo S\'anchez-Rey$^{1}$} 
\author{Guillermo Chac\'on-Acosta$^{2}$}
\author{Leonardo Dagdug$^{3}$}
\author{David Cubero$^{1}$}
\email[]{Corresponding author: dcubero@us.es}

\affiliation{$^{1}$Departamento de F\'{\i}sica Aplicada I, EPS, Universidad 
de Sevilla, Calle Virgen de \'Africa 7, 41011 Sevilla, Spain}
\affiliation{$^{2}$Departamento de Matem\'aticas Aplicadas y Sistemas, Universidad Aut\'onoma Metropolitana-Cuajimalpa,  M\'exico D. F. 01120, M\'exico}
\affiliation{$^{3}$Departamento de F\'{\i}sica, Universidad Aut\'onoma Metropolitana-Iztapalapa, M\'exico D. F. 09340, M\'exico}




\begin{abstract}
We report fully relativistic molecular-dynamics simulations that verify the appearance of thermal equilibrium of a classical gas inside a uniformly accelerated container. The numerical experiments confirm that the local momentum distribution in this system is very well approximated by the J\"uttner function -- originally derived for a flat spacetime -- via the Tolman-Ehrenfest effect. Moreover, it is shown that when the acceleration or the container size is large enough,
the global momentum distribution can be described by the so-called modified J\"uttner function, which was initially proposed as an alternative to the J\"uttner function.
\end{abstract}


\maketitle

\section{Introduction}

Ever since Albert Einstein started formulating its theory of general relativity in 1907 \cite{1907ei,1911ei,1916ei}, the uniformly accelerated system has proven to be a central paradigm of relativity, providing a very simple description of the principle of equivalence. Turned into a vivid thought experiment by Einstein himself \cite{1920ei,1938eiin}, the {\em Einstein's elevator}, it famously led him to the prediction of the bending of light by gravity \cite{1911ei} in the earliest stages in the development of the theory of general relativity, being experimentally confirmed by Eddington in 1919 \cite{2007ken}. Even today, it is one of the most utilized models for explaining the main ideas of general relativity. 

To put more examples, the uniformly accelerated system was used by Hawking and Unruh in the 70s to show how gravitational forces can thermalize quantum fields \cite{1975ha,1976un}, or more rencently, in the derivation of the Planck spectrum of thermal scalar radiation \cite{bo10}, or to explore the relationship between the entropy of a gas and the horizon area as in black holes mechanics \cite{kopa08}.

In this paper, we study the thermal equilibrium of a classical gas inside Einstein's elevator by using molecular dynamics one-dimensional simulations. The same kind of simulations were used in \cite{2007cuetal} to show that the J\"uttner function \cite{1911ju} is the correct generalization of the Maxwell-Boltzmann velocity distribution in special relativity. In Ref.~\cite{2009cudu} it was further shown that the modified J\"uttner function \cite{2007dutaha_2}, originally proposed as an alternative to the J\"uttner function, provides a good description of the numerically measured distribution in special relativity when a parameterization on the particle's proper time is used, which could be useful in situations where decay processes are important, since particle's
life-time issues are usually dealt in terms of proper-time intervals. These numerical experiments, having successfully been used to probe notions such as thermal equilibrium and ergodicity in special relativity, still remain the simplest {\em fully} relativistic molecular dynamics technique, since simulations in higher dimensions \cite{2006alromo,2009duhahi}, though approximately correct for dilute gases, usually assume superluminal interactions during the collision events. A similar difficulty is found generally in relativity for Hamiltonian Mechanics: as shown by the "no-interaction" theorem proved by Currie, Jordan and Sudarshan \cite{1963cujosu}, the Hamiltonian formalism can only be applied to systems constituted by noninteracting particles. In Ref. \cite{L-M}, the generalized J\"uttner distribution function of a noninteracting gas in a uniformly accelerated frame was derived using the Hamiltonian formalism discussed in \cite{moller} and the usual probabilistic assumptions of statistical mechanics for thermal equilibrium. Since interactions are actually necessary to be able to reach equilibrium,  we believe that the numerical tools used in \cite{2007cuetal,2009cudu} -- where particles' interactions drive the system naturally to equilibrium -- are pertinent to verify that the usual ideas of thermal equilibrium can indeed also be applied in a curved spacetime.

The manuscript is organized as follows. First we discuss some relevant analytical results of thermal equilibrium in stationary gravitational fields, specializing some of these results for a gas inside Einstein's elevator in the following section. Details of the relativistic molecular dynamics simulations are given in Sec.~\ref{sec:md}, and the numerical results presented and discussed in Sec.~\ref{sec:results}. Finally, Sec.~\ref{sec:finale} provides a short summary and conclusions.

\section{Thermal equilibrium in stationary fields}
Consider a material particle which is moving freely under the influence of purely gravitational forces. We will characterize its coordinates in an arbitrary coordinate system $\Sigma$ by $x^\mu=(c t,{\bs x})=(c t,x^1,\ldots,x^d)$, where $d$ is the number of spatial dimensions in the system and $c$ the speed of light in vacuum. Unless explicitly stated, in the following, natural units such that $c=1$ will be assumed. The coordinate system $\Sigma$  may be at rest or accelerated, being all effects of gravitation or inertial forces comprised in the metric tensor $g_{\mu\nu}(x^\mu)$. We will assume that the gravitational field created by the material particle is very small compared with external gravitational or inertial forces, so we can regard $g_{\mu\nu}$ as independent of the state of the particle. According to the principle of equivalence, the equation of motion for the freely falling particle is (see for example \cite{weinberg})
\begin{equation}
\frac{du^\mu}{d\tau}+\Gamma^\mu_{\alpha\beta}u^\alpha u^\beta=0,
\label{eq:eqmoU}
\end{equation}
where $u^\mu=dx^\mu/d\tau$ is the particle four-velocity, $\tau$ is the proper time
\begin{equation}
d\tau^2=g_{\alpha\beta}dx^\alpha dx^\beta,
\label{eq:tau}
\end{equation}
and the Christoffel symbols
\begin{equation}
\Gamma^\mu_{\alpha\beta}=\frac{1}{2}g^{\mu\nu}\left(\frac{\partial g_{\nu\alpha}}{\partial x^\beta}+\frac{\partial g_{\nu\beta}}{\partial x^\alpha}-\frac{\partial g_{\alpha\beta}}{\partial x^\nu}\right).
\label{eq:chris}
\end{equation}
The energy-momentum four-vector of the particle in the coordinate system is defined as $p^\mu=(p^0,{\bs p})=m u^\mu$, where $m$ is the particle rest mass. 

Quite importantly, the following equations can be readily derived \cite{moller} from (\ref{eq:eqmoU}) and (\ref{eq:chris}) for the covariant components of the four-momentum $p_\mu=g_{\mu\nu}p^\nu$,
\begin{equation}
\frac{dp_\mu}{d\tau}-\frac{1}{2}\frac{\partial g_{\alpha\beta}}{\partial x^\mu}u^\alpha p^\beta=0.
\end{equation}
If the metric tensor $g_{\mu\nu}$ does not depend explicitly on time $t=x^0$, that is, the gravitational field is {\em stationary}, the energy $p_0=g_{0\nu}p^\nu$ is a constant of motion, 
\begin{equation}
\frac{d p_0}{dt}=0.
\label{eq:p0const}
\end{equation}
Therefore, we may regard $p_0$ as the {\em total} energy of the particle \cite{moller}, "total" because it includes the interaction with the gravitational fields (or with the inertial forces if the coordinate system is accelerating). From the definition of the four-velocity $u^\mu$, it is clear that $u_\mu u^\mu=1$, implying $p_\mu p^\mu=m^2$. This last equation can be used to find $p^0$ as a function of ${\bs p}=(p^1,\ldots,p^d)$, yielding an explicit expression for the total energy $p_0$ in terms of ${\bs x}$ and ${\bs p}$,
\begin{equation}
p_0=\sqrt{(g_{0i}p^i)^2+g_{00}(m^2-g_{ij}p^ip^j)},
\label{eq:p0:explicit}
\end{equation}
where the sum over the latin indices run over the spatial coordinates only, $i,j=1,\ldots,d$.

\paragraph*{Generalized J\"uttner distribution.} 
The particle's total energy (\ref{eq:p0:explicit}) determines the shape of the momentum distribution of a gas in thermal equilibrium under stationary gravitational fields. To show this, let us consider a dilute gas of relativistic particles which is confined to some container at rest in the coordinate system. If we assume that each collision between any two gas particles occurs around an space-time event in which the metric tensor $g_{\mu\nu}$ does not appreciably vary, then the conservation of the energy-momentum $p^\mu_l+p^\mu_n$ of the two particles in the (elastic) collision, together with the linear dependence of $p_0$ on $p^\mu$, $p_0=g_{0\mu}p^\mu$, guarantees that the sum of the total energies $p_{0,l}+p_{0,n}$ is also a collisional invariant. Therefore, the total energy of the system 
\begin{equation}
E_t=\sum_{n=1}^N p_{0,n}
\end{equation}
is conserved throughout the system, provided that there is no energy exchange at the boundaries.
From this point, it is easy to show for a dilute gas, using a similar kinetic theory derivation than in the case of the Boltzmann equation for special relativity \cite{1980dg}, that the one-particle phase-space distribution, defined as
\begin{equation}\label{eq:def_f}
f({\bs x},{\bs p},t)=\frac{1}{N}\sum_{i=1}^N\delta({\bs x}-{\bs x}_i(t))\delta({\bs p}-{\bs p}_i(t)),
\end{equation}
must tend in the proper coordinate system $\Sigma$ where the gas is at rest, for sufficiently large times and large number of particles $N$, to the equilibrium distribution 
\begin{eqnarray}
f({\bs x},{\bs p})&=&Z^{-1}\exp(-\beta p_0)  \label{eq:juttner} \\
&=& Z^{-1}\exp\left[-\beta\sqrt{(g_{0i}p^i)^2+g_{00}(m^2-g_{ij}p^ip^j)}\right], \nonumber
\end{eqnarray}
where $Z$ is a normalization constant given by the condition $\int d^d{\bs x}d^d{\bs p}f({\bs x},{\bs p})=1$, and $\beta=1/(k_B T)$, where $k_B$ is the Boltzmann constant and $T$ the global temperature of the gas (see the discussion below). Eq. (\ref{eq:juttner}) can be regarded as the generalization of the J\"uttner distribution to stationary gravitational fields or inertial forces. For an ideal gas, it contains all the information required to compute any thermodynamic variable.  


\paragraph*{Tolman-Ehrenfest effect.}In the following, let us restrict ourselves to situations where $g_{0i}=0$ for $i=1,\ldots,d$, that is, to the so-called \cite{moller,1934tolman} case of {\em static} fields. Then, assuming also a Cartesian coordinate system in space, $g_{ij}=-\delta_{ij}$ for $i,j=1,\ldots,d$, we can write Eq.~(\ref{eq:juttner}) as
\begin{equation}
f({\bs x},{\bs p})= Z^{-1}\exp[-\beta \sqrt{g_{00}}\sqrt{m^2+p^2}].
\label{eq:juttner2}
\end{equation}
Comparing (\ref{eq:juttner2}) with the original J\"uttner distribution in special relativity 
\cite{1911ju,2007cuetal}
\begin{equation}
f_J({\bs x},{\bs p})= Z_J^{-1}\exp[-\beta_0\sqrt{m^2+p^2}],
\label{eq:juttner0}
\end{equation}
with $\beta_0=1/(k_B T_0)$, yields the Tolman-Ehrenfest equation \cite{1930to,1930toeh}
\begin{equation}
T_0({\bs x})\sqrt{g_{00}({\bs x})}=T,
\label{eq:tolman}
\end{equation}
where $T_0$ is the local temperature at ${\bs x}$. This definition of $T_0$ deserves further discussion. Unlike the temperature transformations between different observers of the traditional relativistic thermodynamics  of Planck and Einstein \cite{1908pl,moller,1934tolman} or Ott \cite{1963ott}, which are based on questionable definitions of 
global quantities \cite{2009duhahi}, the equation (\ref{eq:tolman}) can be provided with distinct physical content. 

Let us consider a second gas confined to a very small container $\mathcal{C}_0$ placed at ${\bs x}$, the container being so small than the metric tensor $g_{\mu\nu}$ does not appreciably vary throughout it. If we allow the gas inside to interact with the gas spread in $\Sigma$, 
eventually both systems will reach thermal equilibrium with a common temperature $T$. 
Then, despite being in an accelerated frame with a possibly non-negligible inertial force -- the container $\mathcal{C}_0$ is at rest in $\Sigma$ --, the momentum distribution function of the gas inside $\mathcal{C}_0$ will be given by the special relativity expression (\ref{eq:juttner0}), being practically indistinguishable from (\ref{eq:juttner2}) in the scale of the container (apart from a trivial normalization constant). Thus, all the local thermodynamic quantities in $\mathcal{C}_0$ will be given by the usual special relativity expressions \cite{2010chdamo}, with a local temperature $T_0$ being given by (\ref{eq:tolman}). 
If the container $\mathcal{C}_0$ is sufficiently small, the larger system will be scarcely affected by the smaller one. Then, the temperature $T$ will be unchanged, and we may properly regard the smaller system as a thermometer device.  If a local observer monitoring this thermometer does not have knowledge of the full metric $g_{\mu\nu}({\bs x})$, only the local temperature $T_0$ will be measured.

In addition, note that an inertial reference frame which is momentarily at rest with the grid point ${\bs x}$ in $\Sigma$ will observe the same particles' momenta inside a small region around ${\bs x}$ -- though in general not the same velocities because of the different proper clocks used as their time coordinate. Since the local statistics given by (\ref{eq:juttner2}) in that region is the same than in special relativity, we are entitled to use the statistical thermometer proposed in Ref.~\cite{2007cuetal} for inertial frames, provided we express it in terms of particle momenta. This thermometer has been used in the simulations presented below. 

Nevertheless, it must be noticed that the Tolman-Ehrenfest effect does not imply that the particle dynamics in a small region of $\Sigma$ will be the same than in an inertial frame with a flat metric. As mentioned above, even if the metric tensor is approximately flat in $\mathcal{C}_0$, the particles inside may well feel the gravitational (or inertial) force.
 Rather, the global equilibrium distribution (\ref{eq:juttner2}) shows that the momentum distribution is not largely affected by this {\em static} force (apart from the Tolman-Ehrenfest effect), in a similar fashion to what happens in the non-relativistic limit, where the Maxwell-Boltzmann distribution remains valid under an external gravitational field.


\section{Einstein's elevator}
Let us now consider a gas inside Einstein's elevator. The elevator itself will be modeled as a rigid container that is being uniformly accelerated along the positive direction of the $z$-axis with respect to an inertial reference frame $\tilde{\Sigma}$. More precisely, the bottom of the container -- chosen as the origin of the accelerated coordinate system $\Sigma_e$ -- has a four-vector acceleration $\tilde{a}^{\mu}=d^2\tilde{x}^{\mu}/{d\tau^2}$ with respect to $\tilde{\Sigma}$ given at any time by $\tilde{a}_{\mu}\tilde{a}^{\mu}=-g^2$. This implies that the bottom  of the container has a constant acceleration $g$ in its proper frame, that is, in an inertial frame momentarily at rest at this point \cite{mithwe00}. To describe the rest of the container we can use the grid points of a coordinate system with rigid axes whose origin is moving with the above mentioned uniformly acceleration. It is well-known, \cite{mithwe00,moller}, that such a accelerated system $\Sigma_e$ has the following metric tensor (taking $d=1$) 
\begin{equation}
g_{\mu\nu}=\mbox{diag}\Big[ (1+gz)^2,-1\Big].
\label{eq:metric}
\end{equation}
This coordinate system -- also called the Rindler coordinates -- fails for $x^1=z\le-1/g$, \cite{mithwe00}, but this is of no consequence to us because we have chosen the lower boundary of the container at $z=0$. Let us locate the top at $z=L$. Note that, despite being at rest with the container's bottom in $\Sigma_e$, from the point of view of $\tilde{\Sigma}$ the top is moving with a reduced proper acceleration -- indeed, an accelerometer at $z=L$ would measure $\tilde{a}_{\mu}\tilde{a}^{\mu}=-g^2/(1+gL)^2$. Thus, an observer in the inertial frame $\tilde{\Sigma}$ would notice a continuous shortening of the container's length, which is easily explained in terms of the Lorentz-Fitzgerald contraction \cite{bell87}. Nevertheless, in an inertial frame which is momentarily at rest with the origin of $\Sigma_e$, the whole container will be seen at rest.  

According to the principle of equivalence, the inertial force observed in $\Sigma_e$ is equivalent to an external gravitational field with a line force running in the negative direction of the $z$-axis. This gravitational field -- which is homogeneous only to a first approximation \cite{1911ei} -- is fully determined by the metric tensor  (\ref{eq:metric}).

Inserting this metric tensor in (\ref{eq:tau}) yields $d\tau=dt/\Gamma(z,v)$, where $v=dz/dt$ is the particle's velocity, and
\begin{equation}
\Gamma(z,v)=\frac{1}{\sqrt{(1+gz)^2-v^2}}
\label{eq:lorentz}
\end{equation}
is a generalized Lorentz factor. This factor allows us to write $u^\mu=\Gamma(1,v)$, which upon insertion in (\ref{eq:eqmoU}) leads to the following equation of motion for freely falling particles
\begin{equation}
\frac{dv}{dt}-\frac{2gv^2}{1+gz}+g(1+gz)=0.
\label{eq:motV}
\end{equation}
The solution of (\ref{eq:motV}), for the generic initial conditions $z(0)=z_0$ and $v(0)=v_0$, is
 \begin{equation}
z(t)=\frac{1}{g} \Bigg[ \frac{(1+g z_0)^2}{(1+gz_0)\cosh (gt)-v_0 \sinh(gt)}-1\Bigg],
\label{eq:zt}
\end{equation}
and
 \begin{equation}
v(t)=\frac{(1+g z_0)^2 \Big[ v_0\cosh(gt)-(1+gz_0)\sinh (gt) \Big]}
{\Big[(1+gz_0)\cosh (gt)-v_0 \sinh(gt)\Big]^2}.
\label{eq:vt}
\end{equation}

The particle's energy is given by $p^0=\sqrt{m^2+p^2}/(1+gz)$. However, we have already seen that the relevant energy in the presence of gravitational or inertial forces is the particle's total energy $p_0$ (\ref{eq:p0:explicit}), shaping the generalized J\"uttner distribution (\ref{eq:juttner2}), and here taking the form
\begin{equation}
f(z,p)=Z^{-1} \exp{\left[-\beta (1+gz)\sqrt{m^2+p^2}\right]},
\label{eq:juttner3}
\end{equation}
with the normalization constant given by
\begin{equation}
Z=\frac{2}{\beta g}\Big[K_0 (\beta m)-K_0\Big(\beta m(1+gL)\Big) \Big]\,,
\end{equation}
and $K_n$ being the modified Bessel functions of the second kind \cite{abst72}.

Note that despite the spatial inhomogeneity introduced by the system acceleration, the distribution (\ref{eq:juttner3}) -- or the more general expression (\ref{eq:juttner2}) -- is symmetric with respect to the momentum $p$, like in the non-relativistic limit. Consequently, the equilibrium average momentum at each point $z$ is zero, and every point of the fluid is at rest in the frame $\Sigma_e$.

\section{Relativistic molecular dynamics}
\label{sec:md}
To verify the appearance of thermal equilibrium in Einstein's elevator, we have performed fully relativistic $d=1$ molecular dynamics simulations, similar to those presented in \cite{2007cuetal,2009cudu} for non-accelerated gases. 

In this model, the gas consists of classical point-particles: $N_1$ light particles of rest mass $m_1$, and $N_2$ heavy particles of rest mass $m_2=2 m_1$. Neighboring particles may exchange momentum and
energy in elastic binary collisions, governed by the relativistic energy-momentum conservation laws \cite{2007cuetal}. Taking place in single space-time points, these collisions are unaffected by the curvature of the metric tensor. Interactions with the container's walls are elastic, i.e. $p\rightarrow-p$ in $\Sigma_e$, thus defining the accelerated frame $\Sigma_e$ as the rest frame of the container. 

Performing the simulations in $\Sigma_e$, the main distinctive feature introduced by the accelerated container is the time evolution of all particles in the intervals between collisions, i.e. the equation of motion (\ref{eq:eqmoU}). In the simulations, all particles are moved according to the formulas (\ref{eq:zt})--(\ref{eq:vt}) in those time intervals .

Another difference with the simulations of Refs. \cite{2007cuetal,2009cudu} is that we have considered here semi-penetrable point-particles. Every time two particles meet at a given space-time point, they exchange momentum with probability $p_t$, remaining unaltered with probability $q_t=1-p_t$. In other words, a fraction $q_t$ of the time, with $q_t\ne 0$, two colliding particles just cross each other like if there would be no interaction between them, allowing each particle to diffuse among the entire simulation box. As it will be discussed later, this rule is aimed at avoiding configurational constraints, highly dependent on the initial conditions, which may inhibit relaxation towards the inhomogeneous equilibrium induced under the influence of the inhomogeneous metric tensor (\ref{eq:metric}).

\section{Numerical results}
\label{sec:results}

\begin{figure}
\includegraphics[width=7cm]{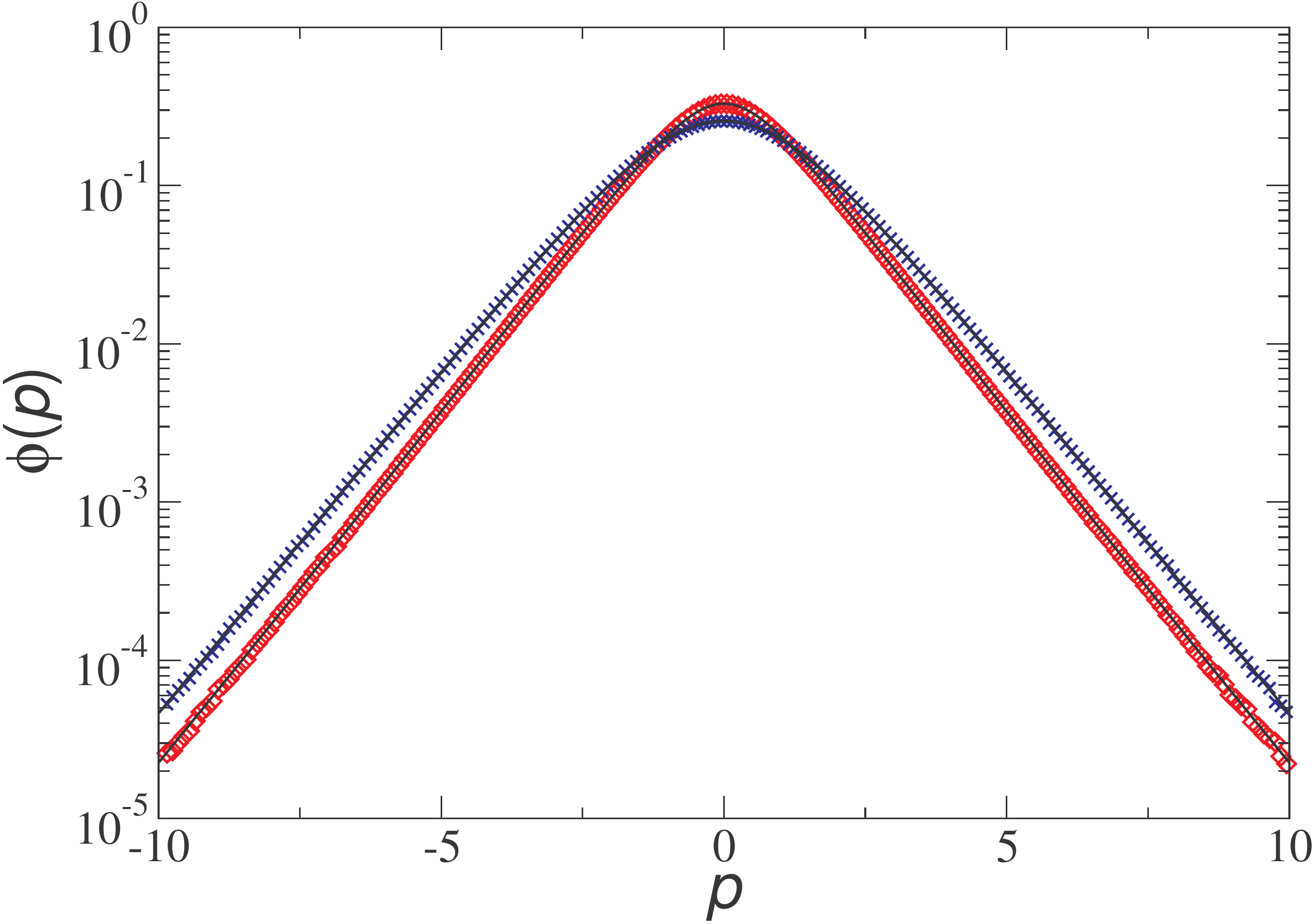}
\caption{(Color online) Numerically measured equilibrium distributions of momenta in Einstein's elevator. The results are based on simulations of an accelerated rigid container of length $L$ with a mixture gas inside consisting of $N_1=415$ light particles with mass $m_1$ (diamonds) and $N_2=585$ particles with mass $m_2=2m_1$ (crosses), and a transparency probability of $q_t=1/2$. Reduced units are defined such that $L=c=m_1=1$.  In these units, the acceleration of the bottom of the container (located at $z=0$) in its proper inertial frame is $g=0.5$. The solid lines correspond to the prediction (\ref{eq:phi}) with same parameter $\beta=0.914$, showing a very good agreement. \label{fig:1}}
\end{figure}

\begin{figure}
\includegraphics[width=7cm]{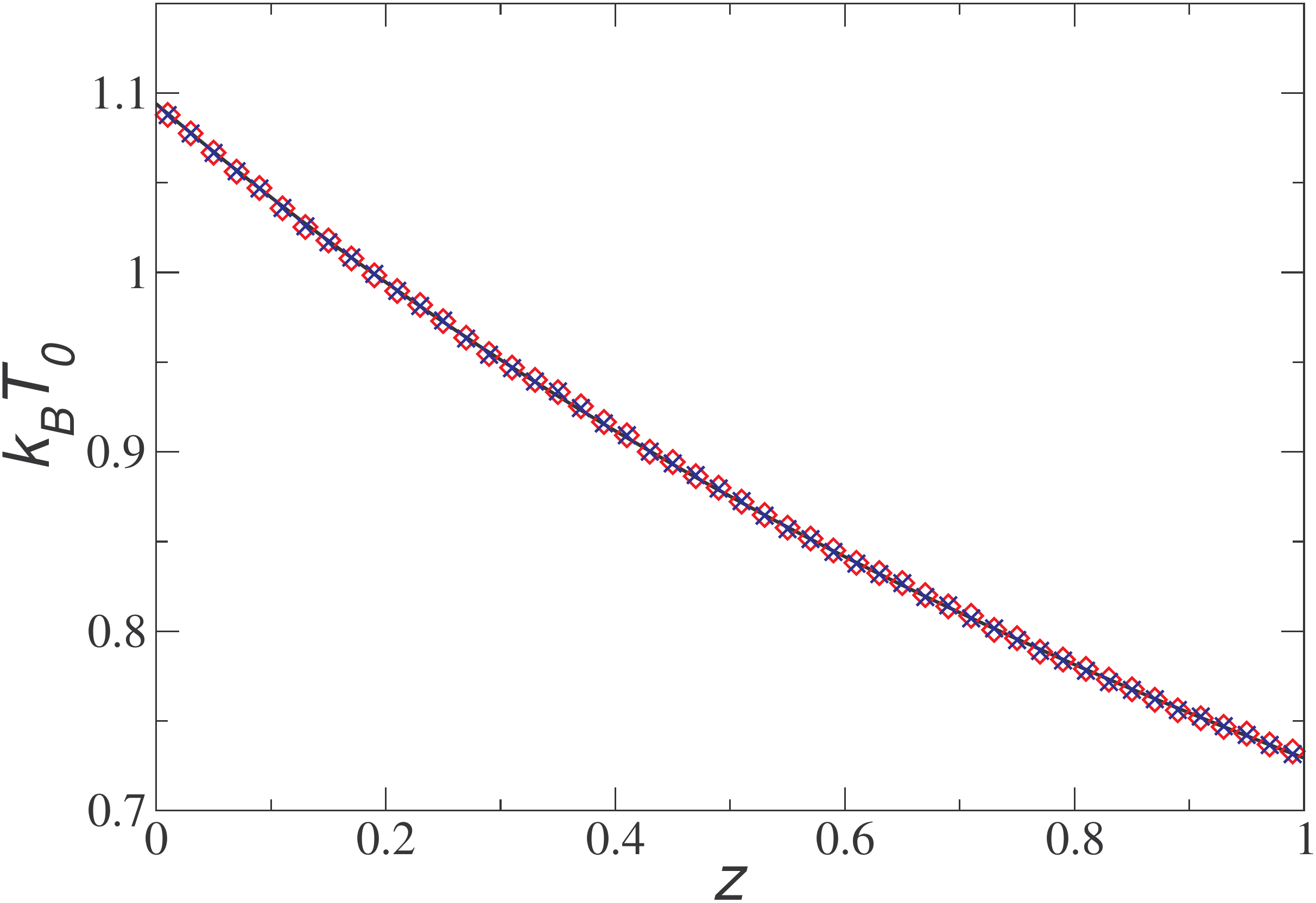}
\caption{(Color online) Tolman-Ehrenfest effect in Einstein's elevator. Local temperature $T_0$ of the light (diamonds) and heavy (crosses) particles as measured locally using the statistical thermometer proposed in Ref.~\cite{2007cuetal}. The solid line is the Tolman-Ehrenfest equation (\ref{eq:tolman2}) with $k_BT=1.094$.  Rest of the parameters as in Fig.~\ref{fig:1}.
\label{fig:2} }
\end{figure}

\begin{figure}
\includegraphics[width=7cm]{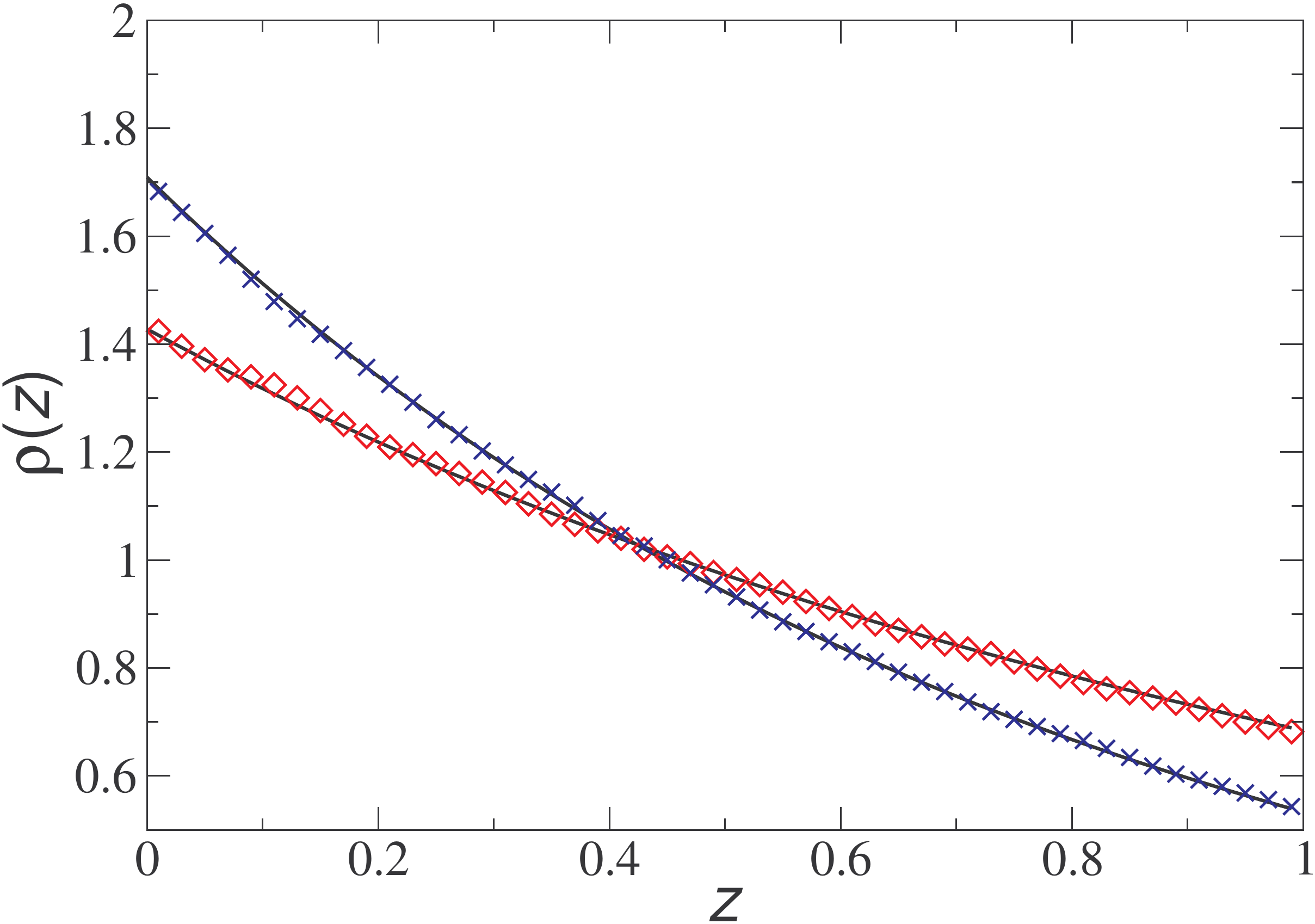}
\caption{(Color online) Particle density in Einstein's elevator for light
particles (diamonds) and heavy particles (crosses). The solid
lines correspond to (\ref{eq:nz}) for both species. Rest of the parameters as in Fig.~\ref{fig:1}.
\label{fig:3}}
\end{figure}

In order to verify the generalized J\"uttner distribution globally, let us start with the marginal distribution of momenta $\phi(p)$, which can be obtained from (\ref{eq:juttner3}) by direct integration
\begin{eqnarray}
\phi(p)= \int_0^L \!\!\! dz\, f(z,p)&=&Z^{-1}e^{-\beta(1+gL/2)\sqrt{m^2+p^2}} \nonumber \\
& &\times \frac{\sinh(\beta g\sqrt{m^2+p^2}/2)}{\beta g\sqrt{m^2+p^2 }/2}.
\label{eq:phi}
\end{eqnarray}
Obviously, in the limit $g\rightarrow 0$ we recover the standard J\"utner momentum distribution \cite{2007cuetal},
\begin{equation}
\phi_\mathrm{J}(p)=  \frac{1}{2m K_1 (\beta m)} \; e^{-\beta
\sqrt{m^2+p^2}}. 
\label{eq:J}
\end{equation}

Figure~\ref{fig:1} depicts the marginal distribution of momenta $\phi(p)$ from one-dimensional simulations as described above for a system with $g=0.5 c^2/L$. In the simulations, reduced units are defined such that $L=c=m_1=1$. Each particle had been given a random initial position and velocity. Once the system reached the equilibrium state, we measured the particle momenta $\Sigma_e$-simultaneously, repeating this procedure many times during a simulation run in order to have a good statistics, and finally collecting the data into a single histogram.  
A very good agreement is found between the analytical prediction (\ref{eq:phi}) and the simulation data for both the light and heavy particles using the parameter $\beta=0.914$. The fact that there is good agreement with the same parameter $\beta=1/k_BT$ is an indication that both gas species have reached a common equilibrium with same global temperature $k_BT=1.094$. 

This global temperature was computed indirectly from the local temperature using the Tolman-Ehrenfest equation (\ref{eq:tolman}), i.e.
\begin{equation}
T=T_0(z)(1+gz).
\label{eq:tolman2}
\end{equation}
The local temperature $T_0(z)$, shown in Fig.~\ref{fig:2}, was measured directly from the simulations  using the statistical thermometer proposed in Ref.~\cite{2007cuetal}. More specifically, we divided the simulation box into 50 bins, and in each bin we measured the local temperature by using the special-relativity formula \cite{2007cuetal}
\begin{equation}
k_B T_0=\langle \frac{p^2}{\sqrt{m^2+p^2}}\rangle,
\label{eq:thermometer}
\end{equation}
where the averages $\langle \cdot\rangle$ are to be computed in an inertial reference frame which is momentarily at rest with the corresponding grid point $z$ of $\Sigma_e$ (since there $\langle p \rangle=0$, see \cite{2007cuetal}). As we have already discussed above, this is equivalent to compute (\ref{eq:thermometer}) in a small region around $z$ in the accelerated frame $\Sigma_e$. Figure~\ref{fig:2} shows a common local temperature of the light and heavy particles, as expected, as well as a very good agreement with the Tolman-Ehrenfest equation (\ref{eq:tolman2}). The value of $T$ used in Figs.~\ref{fig:1} and \ref{fig:2} corresponds to the average over all bins in the system, being the largest fluctuation over the bins smaller than 0.4\% of its magnitude. 

\begin{figure}
\includegraphics[width=7cm]{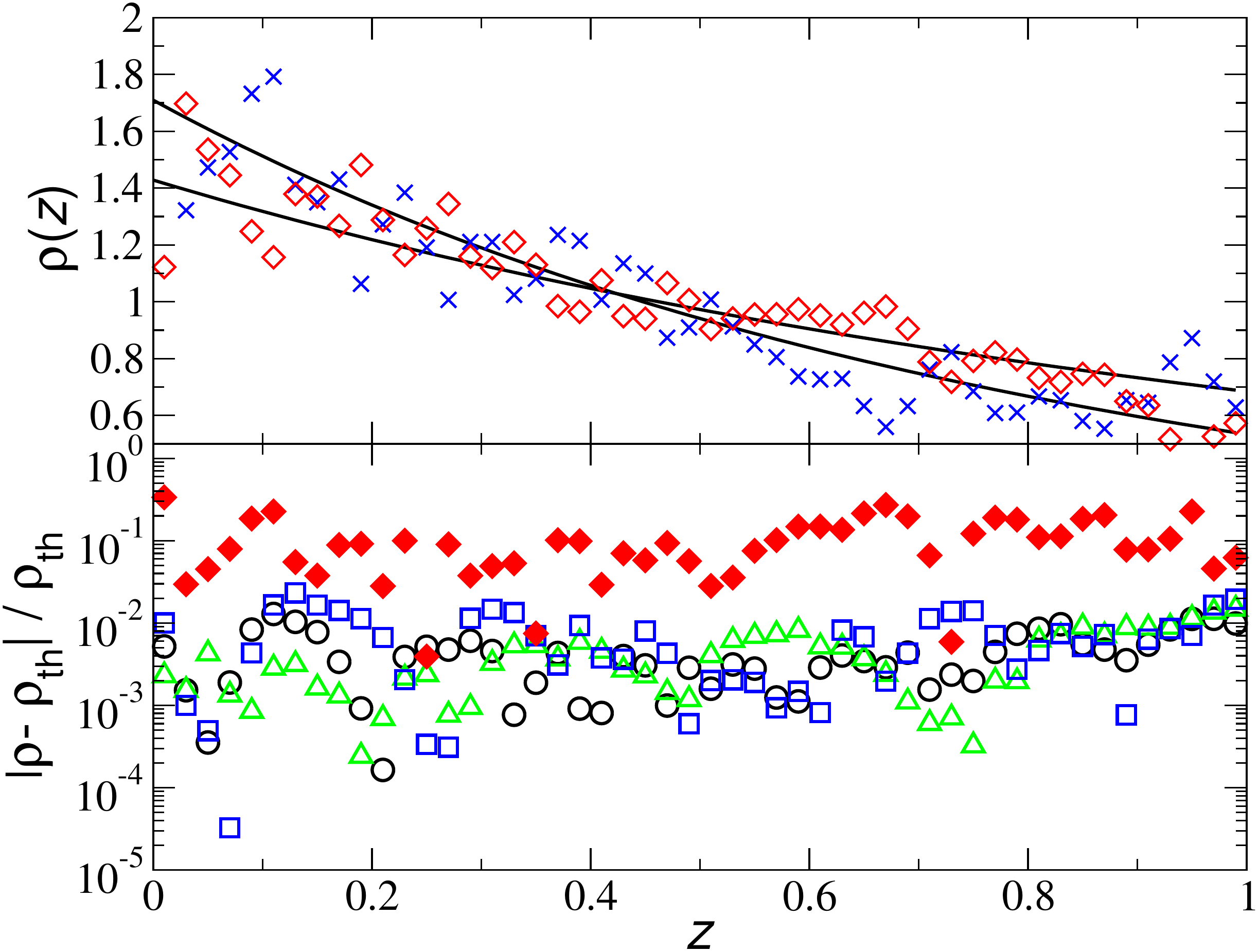}
\caption{(Color online) Sensibility of particle density on the point-particles' semi-penetrability. Top panel shows the same than in Fig. \ref{fig:4} but for a system of impenetrable particles, i.e. $q_t=0$. The bottom panel show the relative error $|\rho-\rho_\mathrm{th}|/\rho_\mathrm{th}$, where $\rho$ is the numerically measured particle density and $\rho_\mathrm{th}$ the analytical prediction given by (\ref{eq:nz}), for the light particles for several transparency probabilities: $q_t=0$ (filled diamonds), $q_t=1/4$ (circles), $q_t=1/2$ (triangles) and $q_t=3/4$ (squares), all simulations starting with the same initial conditions. Rest of the parameters as in Fig.~\ref{fig:1}.
\label{fig:4new}}
\end{figure}

Another macroscopic quantity that can be readily measured in the simulations is the number density $\rho(z)$, which can be easily computed from (\ref{eq:juttner3}) as
\begin{equation}
\rho(z)=\int_{-\infty}^{\infty} \!\!\! dp \; f(z,p) =  \frac{\beta g m 
K_1\left[\beta m(1+gz)\right]}{K_0 [\beta m)]-K_0\left[\beta
m(1+gL)\right]}. 
\label{eq:nz}
\end{equation}
Figure \ref{fig:3} shows a very good agreement between the simulation results and this analytical prediction. 

It is worthwhile to mention that when the particles are not allowed to cross each other, i.e. the case of impenetrable particles  ($q_t=0$), the measured particle density data, though following the same trend, is not so smooth as in Fig. \ref{fig:3}, as shown in the top panel of Fig. \ref{fig:4new}. This is not unexpected, as the geometrical constraint imposed by the spatial one-dimension inhibits the complete relaxation to equilibrium. To understand this fact let us consider a system of impenetrable particles with an initial condition in which light and heavy particles are placed strictly consecutively in the line. After a transient, the particles of one species can accumulate around some point in space so that the local density would approach the corresponding equilibrium value. However, since particles of different species will remain arranged consecutively at any time, there will be the same number of particles of each species inside a small interval around that point, and consequently, the system cannot produce the density differences associated to the thermal equilibrium (\ref{eq:nz}) for each species. If the particles are initially arranged at random, as the case shown in the top panel of Fig. \ref{fig:4new}, the initial relative density fluctuations will survive, producing a poor agreement with the analytical prediction. On the other hand, this effect is expected to disappear when the particles are allowed to diffuse throughout the system with a non-vanishing transparency probability $q_t\ne 0$. This is indeed shown in the bottom panel of Fig. \ref{fig:4new}, where different values of $q_t$ are shown to produce equivalent data, displaying deviations from the analytical prediction that are about an order of magnitude smaller than the data for impenetrable particles. Obviously, the case $q_t=1$ must also be excluded since then there are no collisions to drive the system to equilibrium.  For values of $q_t$ which are very close to the endpoints of the interval $(0,1)$, the behavior will depend on the time scale of the simulation, whether it is large enough so each particle can diffuse throughout the system or there are enough collisions for equilibrium. Finally, let us mention that other quantities such as the global momentum distribution show a better agreement with the analytical prediction in the case $q_t=0$, similar to the observed behavior for other values of $q_t$.
\begin{figure}
\begin{center}
\includegraphics[width=8cm]{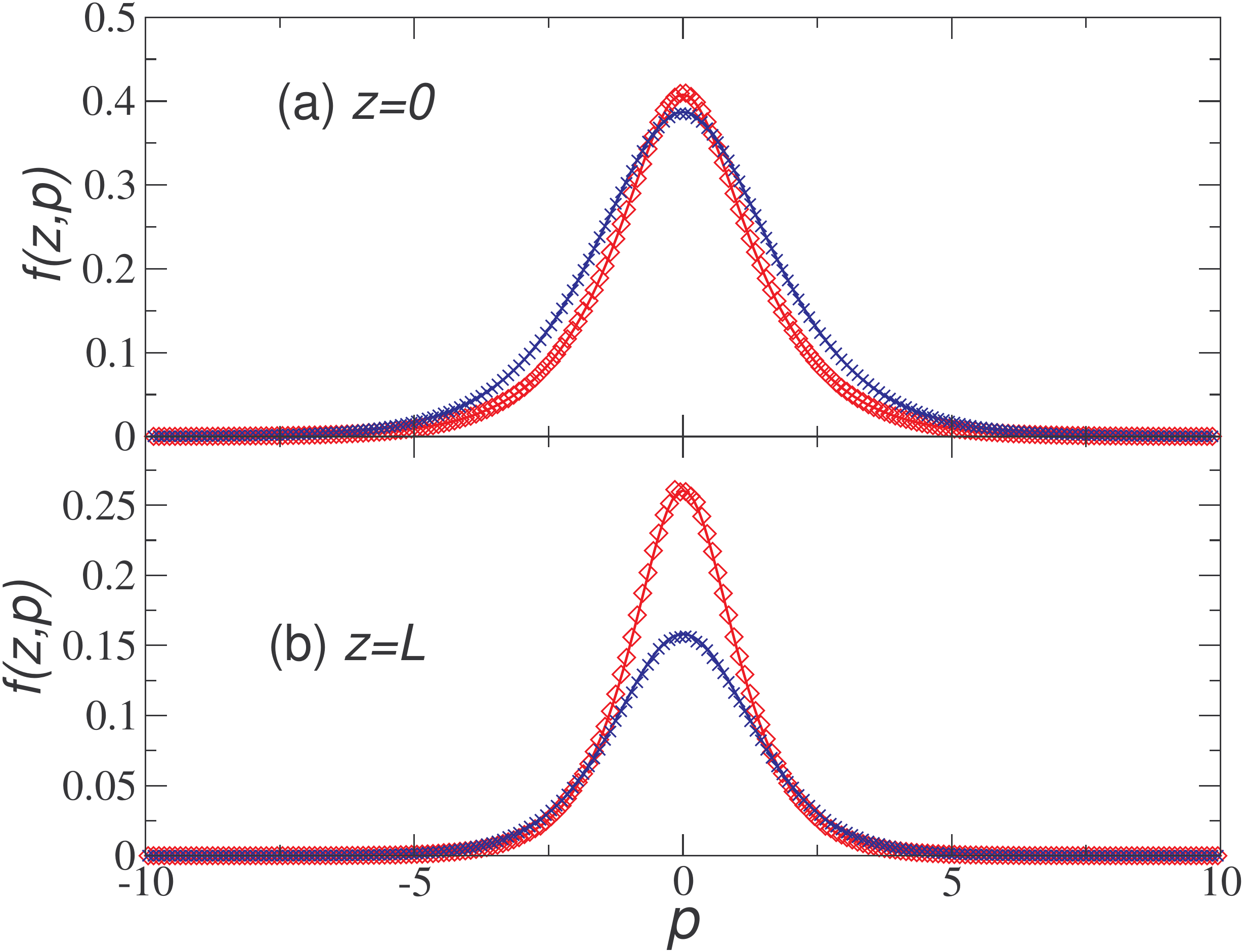}
\caption{(Color online) Momentum distributions $f(z,p)$ at the borders of the
container. The simulation box was divided into 50 bins among the $z$-direction: (a) lower bin, (b) upper bin.
The diamonds (crosses) correspond to light (heavy) particles, and the solid lines to the generalized J\"uttner function (\ref{eq:juttner3}). Rest of the parameters as in Fig.~\ref{fig:1}.
\label{fig:4}}
\end{center}
\end{figure}

So far we have tested global or single-average local macroscopic quantities such as $\phi(p)$ or $T_0(z)$ and $\rho(z)$. A slightly more demanding test can be observed in Fig.~\ref{fig:4}, where the one-particle phase-space distributions at the borders of the system are plotted. Again, an excellent agreement is observed for both species between the simulation results and the analytical prediction -- in this case given by the generalized J\"uttner function (\ref{eq:juttner3}).

\begin{figure}
\includegraphics[width=8cm]{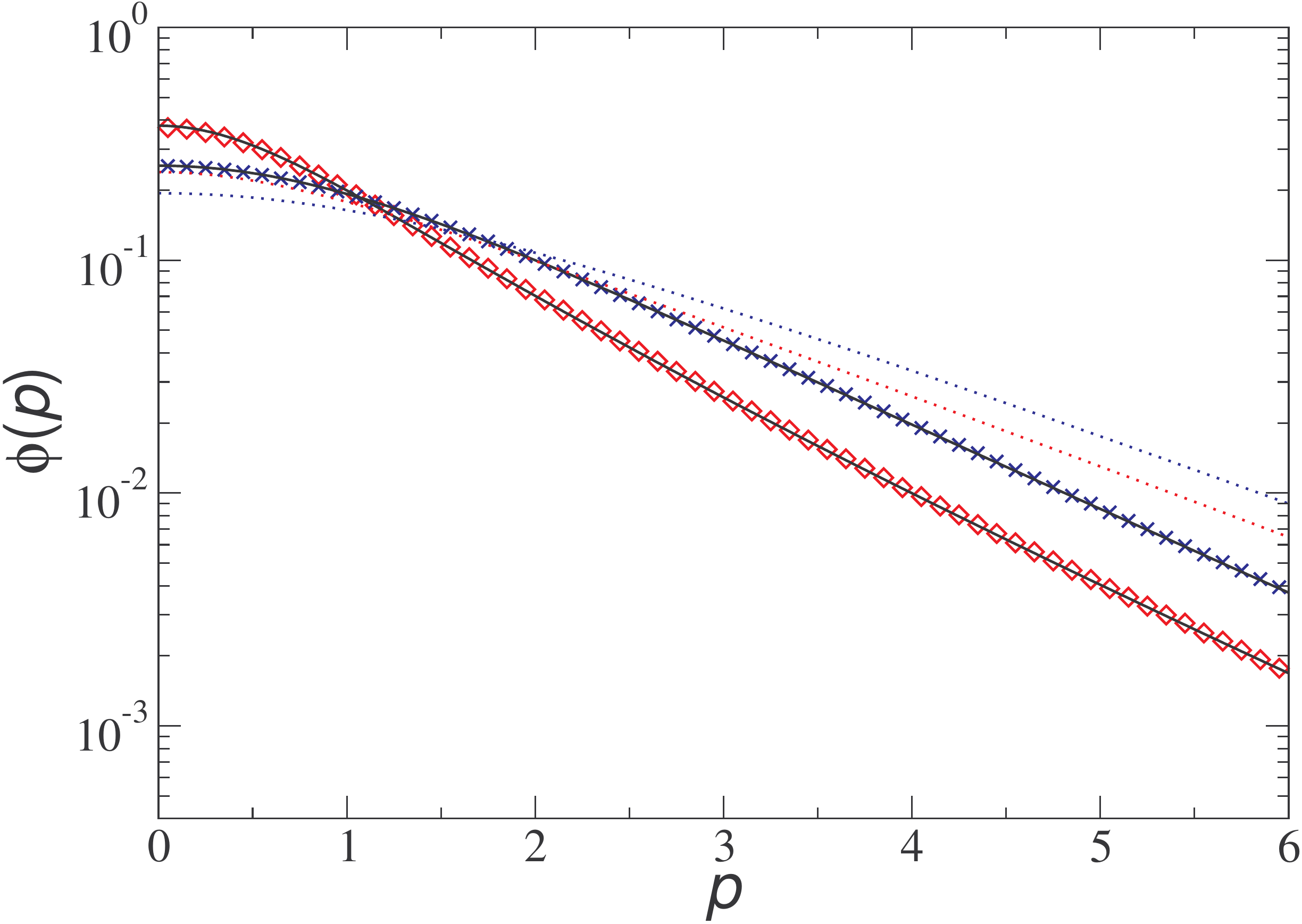}
\caption{(Color online) Equilibrium distributions of momenta in Einstein's elevator for an acceleration $g=5$ (rest of the parameters as in Fig.~\ref{fig:1}). The diamonds (crosses) correspond to the light (heavy) particles. The marginal distributions are very well approximated by the modified J\"uttner function (\ref{eq:MJ}) with $\beta=0.71$ (solid lines). This inverse temperature value was obtained by using the same procedure as in Fig.~\ref{fig:2}. The dotted lines show the special-relativity J\"uttner functions  (\ref{eq:J}) with the same $\beta$, plotted as a reference. As the distributions are symmetric with respect to the origin, only the positive momentum axis is shown.
\label{fig:5}}
\end{figure}

Finally, let us highlight a curious feature that happens when the acceleration $g$, or equivalently the container size $L$, is large enough. Formally, it is easy to check that in the limiting case $gL\rightarrow\infty$ the momentum distribution function (\ref{eq:phi}) tends asymptotically to the modified J\"uttner distribution \cite{2007dutaha_2,2009cudu}
\begin{equation}
\phi_\mathrm{MJ}(p)=  \frac{1}{2 K_0 (\beta m)} \; \frac{e^{-\beta
\sqrt{m^2+p^2}}} {\sqrt{m^2+p^2}}. \label{eq:MJ}
\end{equation}
Therefore, when the dimensionless quantity $gL/c^2$ is large enough, the modified J\"uttner distribution (\ref{eq:MJ}) can be used to approximate the marginal distribution of momenta, as shown indeed in Fig.~\ref{fig:5} for a system with $gL/c^2=5$. Even though this value is not much larger than unity, an excellent agreement between the modified J\"uttner and the measured distribution is observed. In this situation, the gravitational field pushes the particles towards the bottom of the container so strongly that the particle density of both species at $z=L$ is negligible, and thus the top boundary becomes irrelevant. 

This phenomenon is not restricted to the one-dimensional situation considered here. From (\ref{eq:juttner2}), it is easy to show that in a generic system of dimension $d$, the marginal equilibrium distribution in Einstein's elevator is also well approximated by the modified J\"uttner distribution if the acceleration or the system size are large enough.

\section{Summary}
\label{sec:finale}
After discussing some analytical results of the thermal equilibrium of an ideal gas in a stationary gravitational field, leading to the generalized J\"uttner distribution, we have presented numerical evidence of the existence of this thermal equilibrium in a one-dimensional gas confined to a container at rest in a uniformly accelerated system. The numerical results, based on fully relativistic molecular dynamics simulations, also verify the Tolman-Ehrenfest effect for this static system by using the statistical thermometer proposed in Ref.~\cite{2007cuetal} for inertial frames. 

In addition, we have shown that when the acceleration $g$ or the container size $L$ in the direction of the acceleration is large enough so that the upper wall is not needed for confinement, because its role is played by gravitational force, the marginal distribution of momentum in the container becomes the so-called modified J\"uttner function (\ref{eq:MJ}). This fact, together with the observation that the momentum distribution function can always be approximated locally by the standard J\"uttner distribution (\ref{eq:juttner0}), shows that Einstein's elevator is a versatile system where the J\"uttner and modified-J\"uttner functions may refer to different aspects of the same equilibrium distribution.

\begin{acknowledgments}
This research was funded by the Ministerio  de Ciencia e Innovaci\'on of Spain FIS2008-02873 (BSR and DC) and PROMEP 47510283, CONACyT No. 167563 (GCA).
\end{acknowledgments}

%

\end{document}